\input{aipcheck}

\documentclass[
    ,final            % use final for the camera ready runs
  ]
  {aipproc}

\layoutstyle{6x9}

\begin{document}

\title{Energy scale of the Big-Bounce}


\author{W{\l}odzimierz Piechocki}{
  address={Theoretical Physics
Department, Institute for Nuclear Studies,
\\ Ho\.{z}a 69, 00-681 Warszawa, Poland}
}

\begin{abstract}
We examine the nature of the cosmological big-bounce (BB)
transition within the loop geometry underlying Loop Quantum
Cosmology (LQC)  at classical and quantum levels. Our canonical
quantization method is an alternative to the standard LQC.  Our
method opens the door for analyzes  of spectra of physical
observables like  energy density and  volume operator. We find
that one cannot determine the energy scale specific to BB by
making use of the loop geometry without an extra input from
observational cosmology.
\end{abstract}

\maketitle

%%%%%%%%%%%%%%%%%%%%%%%%%%%%%%%%%%%%%%%%%%%%
%% MAINMATTER
%%%%%%%%%%%%%%%%%%%%%%%%%%%%%%%%%%%%%%%%%%%%

\section{Introduction}

Observational cosmology strongly suggests that the universe
emerged from a state with extremely high energy densities of
physical fields. Mathematical cosmology gives numerous examples of
solutions with an initial Big-Bang singularity. This is why it is
commonly believed that the cosmological singularity is a real
feature of the Universe.

It seems that one can treat the problem of the singularity by
making use of loop geometry: classical Big-Bang turns into quantum
Big-Bounce  transition. Thus, one may try to answer some
interesting questions: What is the  energy scale of the
unification of general relativity (GR)  with quantum physics? What
is the structure of spacetime at semi-classical level (simply
connected, foamy or discrete)? What is the  origin of matter,
non-gravitational fields and spacetime? What was  before the
Big-Bounce?

We try to address the first two questions in the context of the
flat FRW universe with massless scalar field.

There are two methods of quantization of the cosmology models
based on loop geometry: (i) standard loop quantum cosmology (LQC),
based on the rule: `first quantize, then impose constraints
\cite{abl,boj}, and (ii) non-standard LQC using the rule: `first
solve constraints, then quantize \cite{ppw1,ppw2}. In the former
case, one believes that the classical Big-Bang is replaced by
quantum Big-Bounce due to strong  quantum effects at the Planck
scale \cite{aps1,aps2,aps3}. In the latter case, one shows that it
is the modification of GR by loop geometry which  is responsible
for the resolution of the singularity and the energy density of
matter at the Big-Bounce is unknown \cite{ppw3}.

\section{Classical Level}

The gravitational part of the GR Hamiltonian reads \cite{ppw1}
\begin{equation}\label{ham1}
    H_g:= \int_\Sigma d^3 x (N^i C_i + N^a C_a + N C),
\end{equation}
where $\Sigma$, space-like part of spacetime $R \times \Sigma$;
$(N^i, N^a, N)$, Lagrange multipliers; $(C_i, C_a, C)$ are Gauss,
diffeomorphism and scalar constraints; $(a,b = 1,2,3)$, spatial
indices; $(i,j,k = 1,2,3)$ internal $SU(2)$ indices.  Constrains
must satisfy the specific algebra. For  flat FRW universe with
massless scalar field (in some special gauge) the Hamiltonian
(\ref{ham1}) turns into \cite{ppw1}
\begin{equation}\label{hamG}
H_g = - \gamma^{-2} \int_{\mathcal V} d^3 x ~N
e^{-1}\varepsilon_{ijk}
 E^{aj}E^{bk}  F^i_{ab}\, ,
\end{equation}
where  $\gamma$, Barbero-Immirzi parameter; $\mathcal V\subset
\Sigma$, elementary cell; $N$, lapse function;
$\varepsilon_{ijk}$, alternating tensor; $E^a_i $, density
weighted triad;  $ F^k_{ab} =
\partial_a A^k_b - \partial_b A^k_a + \epsilon^k_{ij} A^i_a
A^j_b$, curvature of $SU(2)$ connection $A^i_a$; $e:=\sqrt{|\det
E|}$.

Modification of GR by loop geometry means replacement \cite{abl}
of $ F^k_{ab}$ by its  approximation
\begin{equation}\label{finite}
 F^k_{ab}(\lambda)=  \lim_{\mu\,\rightarrow
\,\lambda} \Big\{ -2\;Tr\;\Big(\frac{h^{(\mu)}_{ij}-1}{\mu^2
V_o^{2/3}}\Big)\;{\tau^k}\; ^o\omega^i_a  \; ^o\omega^j_a\Big\},
\end{equation}
and we have
\begin{equation}\label{zero}
F^k_{ab}= \lim_{\lambda\,\rightarrow \,0}\, F^k_{ab}(\lambda),
\end{equation}
where $V_0$ is the so-called fiducial volume. Holonomy of the
connection around the square loop  with sides length $\mu
V_0^{1/3}$ is found to be
\begin{equation}\label{box}
   h^{(\mu)}_{ij} = h^{(\mu)}_i
h^{(\mu)}_j (h^{(\mu)}_i)^{-1} (h^{(\mu)}_j)^{-1}
\end{equation}
The holonomy  along straight edge $ ^oe^a_k\partial_a $ of length
$\mu V_0^{1/3}$ (in fundamental, \\j = 1/2, representation of
SU(2) group) reads
\begin{equation}\label{ho1}
h^{(\mu)}_k (c) =  \exp (\tau_{k}\mu c) = \cos (\mu c/2)\;1 +
2\,\sin (\mu c/2)\;\tau_k,
\end{equation}
where $\tau_k = -i \sigma_k/2\;$ ($\sigma_k$ are the Pauli spin
matrices).

Making  use of the so-called Thiemann identity leads finally to
\begin{equation}\label{hamR}
    H_g = \lim_{\lambda\rightarrow \,0}\; H^{(\lambda)}_g ,
\end{equation}
where
\begin{equation}\label{hamL}
H^{(\lambda)}_g = - \frac{sgn(p)}{2\pi G \gamma^3 \lambda^3}
\sum_{ijk}\,N\, \varepsilon^{ijk}\, Tr \Big(h^{(\lambda)}_i
h^{(\lambda)}_j (h^{(\lambda)}_i)^{-1} (h^{(\lambda)}_j)^{-1}
h_k^{(\lambda)}\{(h_k^{(\lambda)})^{-1},V\}\Big),
\end{equation}
and where $V= |p|^{\frac{3}{2}}= a^3 V_0$ is the volume of the
elementary cell $\mathcal{V}$.  Variables $c$ and $ p$ determine
connections $A^k_a$ and  triads $E^a_k$: $A^k_a =
\,^o\omega^k_a\,c\,V_0^{-1/3} \,$ and $\,E^a_k =
\,^oe^a_k\,\sqrt{q_o}\,p\,V_0^{-2/3} $,\\ where $\,c = \gamma
\,\dot{a}\,V_0^{1/3}$ and $\,|p| = a^2\,V_0^{2/3}$, $\{c,p\} = 8
\pi G \gamma /3$.

Total Hamiltonian for FRW universe with a massless scalar field
$\phi$ reads \cite{ppw1}
\begin{equation}\label{ham}
   H = H_g + H_\phi,
\end{equation}
where $H_g$ is defined by (\ref{hamR}) and  $H_\phi = p^2_\phi
|p|^{-\frac{3}{2}}/2$, and where $\phi$ and $p_\phi$ are
elementary variables satisfying $\{\phi,p_\phi\} = 1$. The
relation $ H \approx 0$ defines the  physical phase space.

Making use of (\ref{ho1}), we calculate  (\ref{hamL}) and get the
modified  total Hamiltonian  corresponding to (\ref{ham})
\begin{equation}\label{regH}
  H^{(\lambda)}/N = -\frac{3}{8\pi G \gamma^2}\;\frac{\sin^2(\lambda
\beta)}{\lambda^2}\;v + \frac{p_{\phi}^2}{2\, v},
\end{equation}
where
\begin{equation}\label{re1}
    \beta := \frac{c}{|p|^{1/2}},~~~v := |p|^{3/2}
\end{equation}
are the  canonical variables (of so-called improved scheme). It
should be emphasized that  (\ref{regH}) presents a purely
classical modified Hamiltonian (with no quantum corrections)!

The Poisson bracket for the canonical variables
$(\beta,v,\phi,p_\phi)$ is defined to be
\begin{equation}\label{re2}
    \{\cdot,\cdot\}:= 4\pi G\gamma\;\bigg[ \frac{\partial \cdot}
    {\partial \beta} \frac{\partial \cdot}{\partial v} -
     \frac{\partial \cdot}{\partial v} \frac{\partial \cdot}{\partial \beta}\bigg] +
     \frac{\partial \cdot}{\partial \phi} \frac{\partial \cdot}{\partial p_\phi} -
     \frac{\partial \cdot}{\partial p_\phi} \frac{\partial \cdot}{\partial
     \phi}.
\end{equation}
The dynamics of a canonical variable $\xi$ is
\begin{equation}\label{dyn}
    \dot{\xi} := \{\xi,H^{(\lambda)}\},~~~~~~\xi \in \{\beta,v,\phi,p_\phi\},
\end{equation}
where $\dot{\xi} := d\xi/d\tau$, and where $\tau$ is an evolution
parameter.  Dynamics in   physical phase space,
$\mathcal{F}_{phys}^{(\lambda)}$, is defined by solutions to
(\ref{dyn}) satisfying the condition $H^{(\lambda)}\approx 0$.
Solutions of (\ref{dyn}) ignoring the constraint
$H^{(\lambda)}\approx 0$ are in    kinematical phase space,
$\mathcal{F}_{kin}^{(\lambda)}$.

If the Hamiltonian is a  constraint which may be rewritten as a
product of a simpler constraint and a function on
$\mathcal{F}_{kin}^{(\lambda)}$ with no zeros, the original
dynamics may be reduced to the dynamics with the simpler
constraint \cite{ppw1}. Equation (\ref{regH}) can be rewritten as
\begin{equation}\label{product}
  H^{(\lambda)} = N\,H_0^{(\lambda)}\,\tilde{H}^{(\lambda)}\approx 0,
\end{equation}
where
\begin{equation}\label{defprod}
H_0^{(\lambda)} := \frac{3}{8 \pi G \gamma^2 v} \;\Big(\kappa
\gamma |p_\phi| + v\,\frac{|\sin(\lambda
\beta)|}{\lambda}\Big),~~~~~~ \tilde{H}^{(\lambda)}:= \kappa
\gamma |p_\phi| - v\, \frac{|\sin(\lambda \beta)|}{\lambda},
\end{equation}
where $\kappa^2 \equiv 4\pi G/3$.

\noindent $H_0^{(\lambda)} = 0$  iff $p_\phi =0=\sin(\lambda
\beta)$. In such  case $\tilde{H}^{(\lambda)} =0$, thus
$H^{(\lambda)}$ equals identically zero so there is no dynamics.
We exclude such pathological cases from further considerations and
assume that $H_0^{(\lambda)}\neq 0.$

For functions $f$ and $g$ on $\mathcal{F}_{phys}^{(\lambda)}$ we
have
\begin{equation}\label{funF}
    \dot{f} = \{f, N H_0^{(\lambda)}\tilde{H}^{(\lambda)}\} = \{f, N H_0^{(\lambda)}\}
    \tilde{H}^{(\lambda)} + N H_0^{(\lambda)}\{f,
    \tilde{H}^{(\lambda)}\} = N H_0^{(\lambda)}\{f,\tilde{H}^{(\lambda)}\},
    \end{equation}
\begin{equation}\label{funG}
    \dot{g} = \{g, N H_0^{(\lambda)}\tilde{H}^{(\lambda)}\} =
    N H_0^{(\lambda)}\{g,\tilde{H}^{(\lambda)}\},~~~~~
    \mbox{for}~~~~~\tilde{H}^{(\lambda)}
    \approx 0.
\end{equation}
The relation
\begin{equation}\label{fraC}
    \frac{\dot{f}}{\dot{g}} = \frac{df}{dg} = \frac{N H_0^{(\lambda)}
    \{f,\tilde{H}^{(\lambda)}\}}
    {N H_0^{(\lambda)}\{g,\tilde{H}^{(\lambda)}\}} = \frac{\{f,\tilde{H}^{(\lambda)}\}}
    {\{g,\tilde{H}^{(\lambda)}\}}
    ,~~~~\mbox{as}~~~~H_0^{(\lambda)} \neq 0,
\end{equation}
may be rewritten as \cite{ppw1}
\begin{equation}\label{integ}
    \frac{df}{\{f,\tilde{H}^{(\lambda)}\}} = \frac{dg}
    {\{g,\tilde{H}^{(\lambda)}\}}
\end{equation}
Thus, in the relative dynamics one canonical variable may be used
as an `evolution parameter', and the dynamics is gauge independent
(no dependance on $N$).

Since the relative dynamics is gauge independent, we choose $N =
1/H_0^{(\lambda)}$ to simplify calculations.  Equations of motion
read \cite{ppw1}
\begin{equation}\label{1a}
\dot{\phi} =
\kappa\gamma~\textrm{sgn}(p_{\phi}),~~~~~\dot{p_{\phi}}=0,
\end{equation}
\begin{equation}\label{2a}
\dot{\beta}=  -4\pi G\gamma \;\frac{|\sin(\lambda\,
\beta)|}{\lambda},~~~
  \dot{v} = 4\pi G\gamma v \cos(\lambda\,
  \beta)~\textrm{sgn}(\sin(\lambda\, \beta)),
\end{equation}
\begin{equation}\label{3a}
\tilde{H}^{(\lambda)}  \approx 0.
\end{equation}
Solution of relative dynamics is found to be \cite{ppw1}
\begin{equation}\label{res2}
2 v = \Delta\cosh\Big(3\kappa\,s\,(\phi - \phi_{0})-\ln \Delta
\Big),
\end{equation}
where
$s:=\textrm{sgn}(p_{\phi}),~\Delta:=\kappa\gamma\lambda|p_\phi|$.
Solution for  $\beta$ may be determined from (\ref{3a}) rewritten
as \cite{ppw1}
\begin{equation}\label{res3}
 v |\sin(\lambda \beta)| = \kappa \,\gamma \,\lambda\,|p_\phi| .
\end{equation}
We can see that the variables $ v$ and $ \beta$ are functions of
an evolution parameter $ \phi$.

A function, $\mathcal{O}: \mathcal{F}_{kin}^{(\lambda)}\rightarrow
R$, is a Dirac observable (we choose $N = 1/H_0^{(\lambda)}$ as
our method is gauge invariant) if
\begin{equation}\label{dirac}
\{\mathcal{O},H^{(\lambda)}\}=  \{\mathcal{O},N H_0^{(\lambda)}
\tilde{H}^{(\lambda)}\}= \{\mathcal{O}, \tilde{H}^{(\lambda)}\} =
0.
\end{equation}
Thus $\mathcal{O}$ is solution to the equation \cite{ppw1}
\begin{equation}\label{dir}
\frac{\sin(\lambda\beta)}{\lambda}\,\frac{\partial
\mathcal{O}}{\partial\beta} - v \cos(\lambda\beta)\,\frac{\partial
\mathcal{O}}{\partial v} -
\frac{\kappa\gamma\,\textrm{sgn}(p_{\phi})}{4 \pi
G}\,\frac{\partial \mathcal{O}}{\partial\phi} = 0.
\end{equation}
Solutions to (\ref{dir}) are found to be \cite{ppw1}
\begin{equation}\label{obser1}
\mathcal{O}_1:= p_{\phi},~~~\mathcal{O}_2:= \phi -
\frac{s}{3\kappa}\;\textrm{arth}\big(\cos(\lambda \beta)\big),~~~~
\mathcal{O}_3:= s\,v\, \frac{\sin(\lambda \beta)}{\lambda}.
\end{equation}
Observables satisfy the Lie algebra \cite{ppw1}
\begin{equation}\label{ala1}
\{\mathcal{O}_2,\mathcal{O}_1\}=
1,~~~~\{\mathcal{O}_1,\mathcal{O}_3\}= 0,~~~~
\{\mathcal{O}_2,\mathcal{O}_3\}=  \gamma\kappa .
\end{equation}
Due to the constraint $\tilde{H}^{(\lambda)}=0$, we have
\begin{equation}\label{con}
\mathcal{O}_3=  \gamma \kappa \,\mathcal{O}_1.
\end{equation}
Thus, in the physical phase space,
$\mathcal{F}_{phys}^{(\lambda)}$, we have only two observables
which satisfy the algebra
\begin{equation}\label{alg1}
\{\mathcal{O}_2,\mathcal{O}_1\}= 1.
\end{equation}

Our kinematical phase space, $\mathcal{F}_{kin}^{(\lambda)}$, is
four dimensional. In  relative dynamics one variable is used to
parameterize three others. Since the  constraint relates two
variables, we have only two independent variables. This is the
reason  we have only  two observables.

We consider functions which can be expressed in terms of
observables and an evolution parameter $ \phi$  so they are not
observables. They do become observables for each  fixed value of $
\phi$, since in such case they are only functions of observables.

In what follows we consider the energy density of matter and the
volume function. The energy density of matter field reads
\begin{equation}\label{rho2}
\rho(\lambda,\phi)=\frac{1}{2}\,\frac{p_{\phi}^2}{v^2}.
\end{equation}
In terms of observables we have \cite{ppw1}
\begin{equation}\label{obser}
p_\phi = \mathcal{O}_1,~~~~v = \kappa\gamma\lambda\,
    |\mathcal{O}_1|\,\cosh\big(3\kappa  (\phi- \mathcal{O}_2)\big).
\end{equation}
The density $\rho$ takes its maximum at the minimum of $v$
\begin{equation}\label{cr1}
\rho_{\max} = \frac{1}{2\kappa^2 \gamma^2}\,\frac{1}{ \lambda}^2.
\end{equation}
Let us apply (\ref{cr1}) to the  Planck scale: $l_{Pl}:=
\sqrt{\hbar G/c^3}\sim 10^{-35} m ;~~~\rho_{Pl}:= c^5/\hbar G^2
\sim 10^{19} GeV$.  Equation (\ref{cr1})  in terms of suitable
units reads\footnote{We use $\gamma\simeq 0.24$ determined in the
black hole entropy calculations.}
\begin{equation}\label{ps1}
 \rho_{\max} = \frac{3\; c^2}{8 \pi G
 \gamma^2}\;\frac{1}{\lambda^2}.
\end{equation}
Substituting $ \lambda = l_{Pl}$  into (\ref{ps1}) gives $
\rho_{max}/\rho_{Pl} \simeq 2,07$; for $ \rho_{\max} = \rho_{Pl}$
we get $ \lambda \simeq 1,44\;l_{Pl}$. Surprisingly, the classical
expression (\ref{ps1})  fits the Planck scale.

The volume function reads: $V(\phi) = v$. It results from the
constraint equation
\begin{equation}\label{ress3}
 v\,|\sin(\lambda \beta)|
= \kappa \gamma \lambda |p_\phi|
\end{equation}
that at the  classical level the volume function is  bounded from
below \cite{ppw1}
\begin{equation}\label{mins}
    v_{\min} = \kappa\gamma\lambda\,|p_\phi|.
\end{equation}
For $ \lambda=0$ the minimum  of $ v$ vanishes!  An interesting
question is: Does the quantum volume operator have a non-zero
minimum eigenvalue?

\section{Quantum Level}

Our quantization method of constrained systems is simple enough to
be fully controlled analytically. In the Schr\"{o}dinger
representation (since $\mathcal{O}_1, \mathcal{O}_2 \in R$) we
have
\begin{equation}\label{quant1}
    \mathcal{O}_2 \mapsto \widehat{\mathcal{O}}_2:=x~~,
    ~~\mathcal{O}_1 \mapsto
    \widehat{\mathcal{O}}_1:= -i\hbar\partial_x ,
\end{equation}
where $x\in R$, and where the  carrier space is $L^2(R)$. Thus,
the representation of (\ref{alg1}) reads
\begin{equation}\label{quant2}
   [\widehat{\mathcal{O}}_2,\widehat{\mathcal{O}}_1]=i\hbar .
\end{equation}
The representation is essentially  self-adjoint on
$C^\infty_0(R)\subset L^2(R)$.

In the representation (\ref{quant1}) the energy density operator
reads \cite{ppw2}
\begin{equation}\label{quant3}
    \rho\rightarrow \widehat{\rho}:=\frac{1}
    {2(\kappa\gamma\lambda)^2\cosh^2[3\kappa(\phi-x)]}.
\end{equation}
Solution to the eigenvalue problem
\begin{equation}\label{eigen}
   \widehat{\rho}\psi=\rho(x_0)\psi ,
\end{equation}
for fixed value of $\phi$, is found to be \cite{ppw2}
\begin{equation}\label{quant4}
    \psi (x)=\delta(x-x_0),~~~\rho(x_0) = \frac{1}{2(\kappa\gamma
\lambda)^2}\frac{1}{\cosh^2\big(3\kappa(\phi-x_0)\big)}.
\end{equation}
Energy density has the same functional form as the classical one!
The  range of spectrum is: $(0,\frac{1}{2(\kappa \gamma
\lambda)^2}).$ We can determine $\rho_{\max}$ if we know $
\lambda$. But, $ \lambda$ is a  free parameter of our method!
Thus, finding the critical density of matter corresponding to the
Big-Bounce is an open problem \cite{ppw3}.

Since we have $V=v$, where
\begin{equation}\nonumber
    v = \kappa\gamma\lambda\,
    |\mathcal{O}_1|\,\cosh[3\kappa  (\phi-
    \mathcal{O}_2)] = |\kappa\gamma\lambda\,
    \mathcal{O}_1\,\cosh[3\kappa  (\phi-
    \mathcal{O}_2)]| =: |w|,
\end{equation}
we first examine the spectrum of $\widehat{w}$. The quantum
operator corresponding to $w$ is defined as follows \cite{ppw2}
\begin{eqnarray}\nonumber
    \widehat{w}\,\psi :=\kappa\gamma\lambda\,\frac{1}{2}\,\bigg(
    \widehat{\mathcal{O}}_1\,\cosh[3\kappa  (\phi-
    \widehat{\mathcal{O}}_2)] +\cosh[3\kappa  (\phi-
    \widehat{\mathcal{O}}_2)]\widehat{\mathcal{O}}_1\bigg)\\
    = i\,\frac{\kappa\gamma\lambda\hbar}{2}\bigg(
    +2 \cosh[3\kappa(\phi-x)]\frac{d}{dx}  -3\kappa\sinh[3\kappa(\phi-x)]\bigg)\,\psi,
\end{eqnarray}
where $\psi \in D(\widehat{w})$. The solution to the eigenvalue
problem of the operator $ \widehat{w}$ reads \cite{ppw3}
\begin{equation}\label{quant6}
    \widehat{w}\psi_b =  b\, \psi_b,~~~~~b \in R,
\end{equation}
\begin{equation}\label{quant7}
  \psi_ b (x):= \frac{\sqrt{\frac{3\kappa}{\pi}}\exp(i \frac{2  b}
 {3\kappa^2
    \gamma\lambda\hbar}\arctan
    e^{3\kappa(\phi-x)})}{\cosh^{\frac{1}{2}}3\kappa(\phi-x)}.
\end{equation}
For the  volume operator, corresponding to $V = v = |w|$, we have
\begin{equation}\label{quant8}
    V \longrightarrow \hat{V} f_b = |\hat{w}| f_b := |b| f_b.
\end{equation}
Thus, the spectrum of the operator $\widehat{V}$ seems to be
continuous, and $ b=0$ is the smallest eigenvalue of $\hat{V}$.
However, it turns out that the spectrum is discrete \cite{ppw2}.

\section{Minimum length problem }

Determination of $\lambda$ by  standard LQC
means \cite{aps2}:
\begin{itemize}
      \item considering eigenvalue problem for the  area operator,
      $\widehat{Ar} = \widehat{|p|}$, in kinematical phase space of LQC:
      $ \widehat{Ar}\,|\mu> = \frac{4\pi \gamma l^2_p}{3}\,|\mu| \,|\mu>
      =: ar (\mu)\,|\mu>,$ where $ar (\mu)$ are continuous since $\mu\in R$;
      \item making reference to  discrete eigenvalues, $\{ 0, \Delta,\ldots\}$,
      of kinematical $\widehat{Ar}$ of  LQG, where $\Delta :=2\sqrt{3}\,\pi\gamma l^2_p$;
      \item assuming: $ ar( \lambda) \equiv \Delta$, which leads to $ \lambda = 3\sqrt{3}/2$.
 \end{itemize}
One assumes in the  standard LQC that  a surface cannot be
squeezed to the zero value due to the existence in the universe of
the minimal quantum of area. Physical  justification for this
assumption, offered by standard LQC, is doubtful because
\cite{ppw4}:
\begin{itemize}
      \item $\widehat{Ar}$ was examined so far only in
      kinematical Hilbert space of LQG, i.e. spectrum of
      $\widehat{Ar}$ ignores the algebra of  constraints of
      LQG  so it has no  physical meaning;
      \item  discrete spectrum of LQG was used to replace
       continuous spectrum of LQC (spectral discretization
      `by hand');
      \item LQC  is not a cosmological sector of LQG,
       but a quantization method  inspired by  LQG. Thus,
       LQC and LQG are models of  different systems.
   \end{itemize}
The assumption of the standard LQC that low-lying eigenvalue of
the area operator of LQG determines the basic parameter of LQC is
{\it ad hoc} \cite{ppw4,boj2}.

\section{ Conclusions}

The modification of the classical Hamiltonian realized by making
use of the loop geometry, parameterized by $ \lambda$, turns
Big-Bang into Big-Bounce. However,  $ \lambda$ is a free parameter
to be determined.

The determination of $\lambda$ is presently a great challenge. It
seems that the FRW model is useless in this context; no eigenvalue
of the volume operator is privileged. An extension of study to
homogeneous (Bianchi I, ...) and isotropic (Lema\^{i}tre, ...)
models  cannot probably change the situation. It seems that it is
the observational cosmology which may bring some resolution to the
problem. So far, the cosmic photons reveal no dispersion \cite{fa}
up to the energy scale  $5\times 10^{17}$ GeV. One hopes that the
primordial gravitational waves have made some imprints on the CMB
spectrum so they may be used to determine the allowed range of
values  for  the parameter $\lambda$.

\begin{theacknowledgments}
The author would like to thank the organizers for inspiring
atmosphere at the Meeting and partial financial support.
\end{theacknowledgments}



\bibliographystyle{aipproc}   % if natbib is available

%\bibliography{sample}

%\newcommand\doingARLO[2][]{%
%  \ifx\mmref\undefined #1\else #2\fi
%}
%
%\doingARLO[\bibliographystyle{aipproc}]
%          {\ifthenelse{\equal{\AIPcitestyleselect}{num}}
%             {\bibliographystyle{arlonum}}
%             {\bibliographystyle{arlobib}}
%          }
%\bibliography{sample}

\end{document}